\documentclass[10pt,letterpaper]{article}
\usepackage[top=0.85in,left=2.75in,footskip=0.75in]{geometry}

\usepackage{changepage}

\usepackage[utf8]{inputenc}

\usepackage{textcomp,marvosym}

\usepackage{fixltx2e}

\usepackage{amsmath,amssymb}

\usepackage{cite}

\usepackage{nameref,hyperref}

\usepackage[right]{lineno}

\usepackage{microtype}
\DisableLigatures[f]{encoding = *, family = * }

\usepackage{rotating}

\raggedright
\usepackage[aboveskip=1pt,labelfont=bf,labelsep=period,justification=raggedright,singlelinecheck=off]{caption}

\makeatletter
\renewcommand{\@biblabel}[1]{\quad#1.}
\makeatother

\date{}

\usepackage{lastpage,fancyhdr,graphicx}
\usepackage{epstopdf}
\fancyheadoffset[L]{2.25in}
\fancyfootoffset[L]{2.25in}

\begin{document}

\begin{flushleft}
{\Large
\textbf{Geometric Mixing, Peristalsis, and the Geometric Phase of the Stomach}
}
\\

Jorge Arrieta$^{1,2}$, 
Julyan H. E. Cartwright$^{3}$, 
Emmanuelle Gouillart$^{4}$, 
Nicolas Piro$^{5\ast}$,
Oreste Piro$^{6}$, 
Idan Tuval$^{1}$
\\

$^{1}$ Mediterranean Institute for Advanced Studies (CSIC-UIB), E-07190 Esporles, Spain
\\
$^{2}$ \'Area de Mec\'anica de Fluidos, Universidad Carlos III de Madrid
\\
$^{3}$ Instituto Andaluz de Ciencias de la Tierra, CSIC--Universidad de Granada,
Campus Fuentenueva, E-18071 Granada, Spain
\\
$^{4}$ Surface du Verre et Interfaces, UMR 125 CNRS/Saint-Gobain, 93303 Aubervilliers, France
\\
$^{5}$ \'Ecole Polytechnique F\'ed\'erale de Lausanne, CH-1015 Lausanne, Switzerland
\\
$^{6}$ Departament de F\'{\i}sica, Universitat de les Illes Balears, E-07071 Palma de Mallorca, Spain
\\

$\ast$ E-mail corresponding author: nicolas.piro@epfl.ch \\Authors are listed in alphabetical order
\end{flushleft}

\section*{Abstract}

Mixing fluid in a container at low Reynolds number --- in an inertialess environment ---
is not a trivial task. Reciprocating motions merely lead to cycles of mixing and unmixing,
so continuous rotation, as used in many technological applications, would appear to be
necessary. However, there is another solution: movement of the walls in a cyclical fashion
to introduce a geometric phase. We show using journal--bearing flow as a model that such
geometric mixing is a general tool for using deformable boundaries that return to the same
position to mix fluid at low Reynolds number. We then simulate a biological example: we
show that mixing in the stomach functions because of the ``belly phase'': peristaltic
movement of the walls in a cyclical fashion introduces a geometric phase that avoids
unmixing.

\section*{Introduction}
How may fluid be mixed at low Reynolds number? Such mixing is normally performed with a
stirrer, a rotating device within the container that produces a complex, chaotic flow.
Alternatively, in the absence of a stirrer, rotation of the container walls themselves can
perform the mixing, as occurs in a cement mixer. On occasions, however, mixing is
attempted by a cyclic deformation of the container walls that does not allow for a net
relative displacement of the corresponding surfaces, situations that often occur both in
artificial devices and in living organisms. At the lowest Reynolds numbers, under what is
known as creeping flow conditions, fluid inertia is negligible, fluid flow is reversible,
and an inversion of the movement of the stirrer or the walls leads --- up to perturbations
owing to particle diffusion --- to unmixing, as Taylor \cite{taylor_film} and Heller
\cite{heller} demonstrated. This would seem to preclude the use of reciprocating motion to
stir fluid at low Reynolds numbers; it would appear to lead to perpetual cycles of mixing
and unmixing. The question then arises of how cyclic changes in the shape of the
containers could lead to efficient mixing. Consider a biological case of cavity flow: the
stomach. In the stomach food and drink are mixed to form a homogeneous fluid termed chyme,
which is then digested by the intestines. Gastric mixing is produced by what is called
peristalsis: by the stomach walls moving in a rhythmic fashion. In mathematical terms, the
shape of the stomach walls undergoes a closed cycle in the space of shapes during each
peristalsis cycle. Obviously only shape cycles that do not require a cumulative net
displacement between any two sections of the stomach can be considered. How then is this
peristaltic movement of the stomach walls able to produce mixing, especially in animals in
which the stomach dimensions are such that fluid inertia of the stomach contents is
negligible?

The solution to this conundrum involves the concept of \emph{geometric phase}. A geometric
phase \cite{shapere} is an example of anholonomy: the failure of system variables to
return to their original values after a closed circuit in the parameters. In this Letter
we propose what we term geometric mixing: the use of the geometric phase introduced by
nonreciprocal cycling of the deformable boundaries of a container as a tool for fluid
mixing at low Reynolds number. To exemplify how this process leads to efficient mixing, we use the well-known two-dimensional mixer based on the journal bearing flow but subject to a much-less-studied
rotation protocol that satisfies the geometrical constraints of cyclic boundary deformations. We lastly show that peristalsis, besides its contribution to important biological functions such as fluid transport within individual tubular organs \cite{Jaffrin1971, Ishikawa2011} or signaling throughout complex biological structures \cite{Alim2013}, fulfills its central role in gastric mixing and digestion
\cite{GItextbook1,GItextbook3,GItextbook2} by operating thanks to a geometric phase in the
stomach.


\section*{Results}
\subsection*{Journal bearing flow}

Taylor \cite{taylor_film} and Heller \cite{heller} used the Couette flow of an
incompressible fluid contained between two concentric cylinders to demonstrate fluid
unmixing due to the time reversibility of the Stokes regime. They showed that after
rotating the cylinders through a certain angle, it is possible to arrive back at the
initial state --- to unmix the flow --- by reversing this rotation through the same angle
with the opposite sign, even when the angle is large enough that a blob of dye placed in
the fluid has been apparently well mixed. Considering as parameters in this device the
positions of the outer and inner cylindrical walls of the container specified respectively
with the angles $\theta_1$ and $\theta_2$ from a given starting point, a geometric phase
might arise from driving this system around a loop in the parameter space.

In a fluid system in the Stokes regime, like ours, as inertia is negligible the motion is
by definition always adiabatic and only induced by the change in the parameters: the
positions of the cylinders. Therefore, any resulting phase after a complete cycle in
parameters is a geometric phase. In the Heller--Taylor demonstration the parameter loop is
very simple: $\theta_1$ first increases a certain amount and then decreases the same
amount while $\theta_2$ remains fixed. This loop encloses no area, and reversibility
ensures that the phase is zero. More complex \emph{zero-area} loops can be constructed by
combining in succession arbitrary pairs of reciprocal rotations of both cylinders, and
they also lead to a null phase. We shall call these constructs \emph{reciprocal cycles}.
In order to consider less trivial loops, we may first note that the parameter space is
homotopic to a 2-torus. Loops on such a space can be classified according to the number of
complete turns that both parameters accumulate along the loop. Note also that a relative
rotation of $2 \pi$ between the walls brings the container to the original configuration
except for a global rotation. Since we are interested in shape loops that can be achieved
without a net cumulative displacement of the surfaces of the containers, we need to
consider only the class of type-0 or \emph{contractible} (to a point) loops.

\begin{figure}[ht]
\begin{center}
\includegraphics[width=1.0\columnwidth]{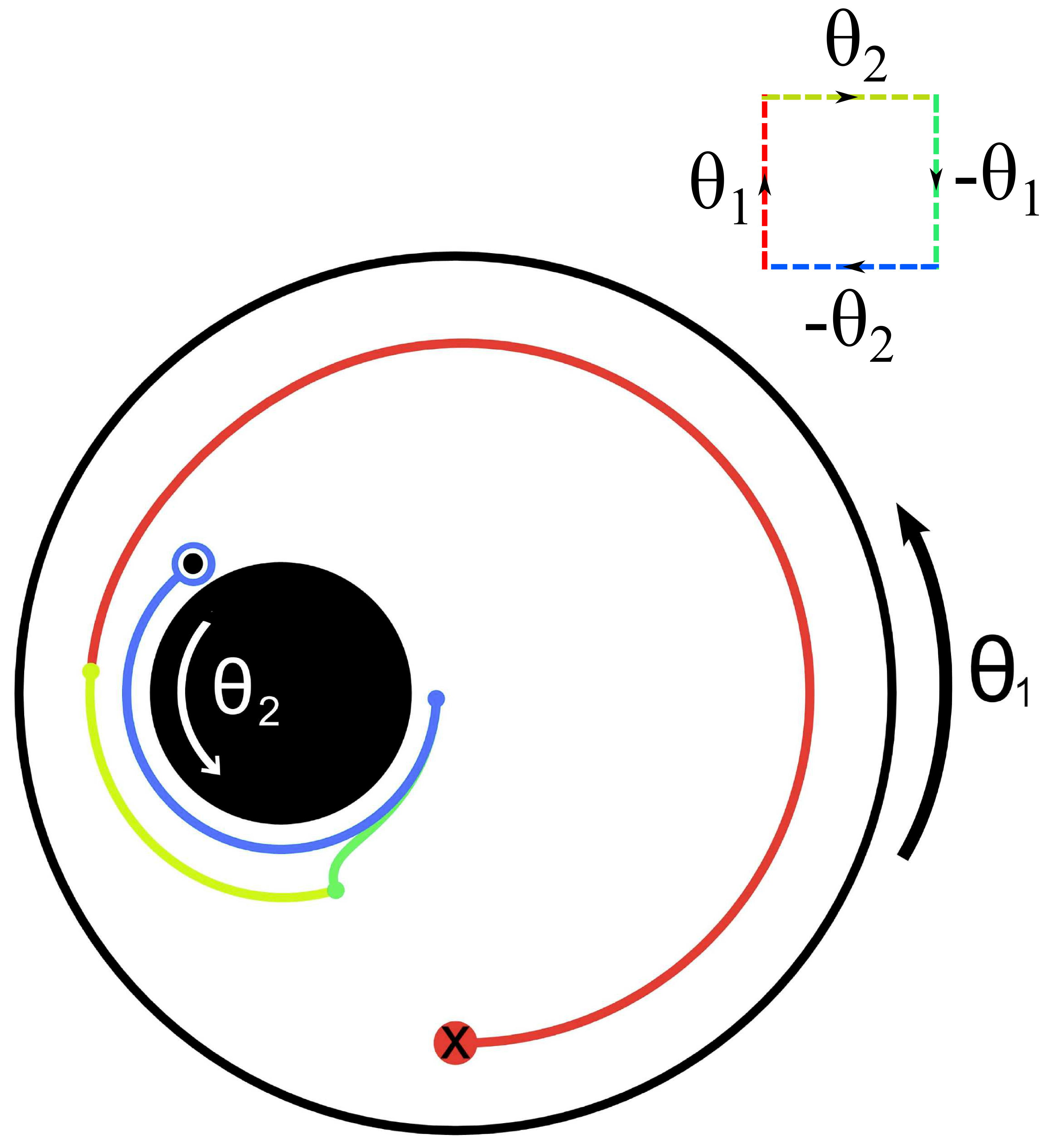}
\end{center}
\caption{{\bf A finite-area non-reciprocal contractible loop.} The journal bearing flow
with cylinder radii $R_1=1.0$, $R_2=0.3$ and eccentricity $\varepsilon=0.4$, taken around
a closed square parameter loop with $\theta_1=\theta_2\equiv\theta=2\pi$ radians. The four
segments of the loop are plotted in different colours (red, yellow, green, blue) to enable
their contributions to the particle motion to be seen. A trajectory beginning at
$(0.0,\,-0.8)$ is shown. The inset shows the performed loop in parameter space.}
\label{jbf_traj}
\end{figure}

All zero-area reciprocal loops are contractible, but there are many more enclosing a
finite area. To obtain a finite-area non-reciprocal contractible loop we can, for
instance, rotate first one cylinder, then the other, then reverse the first, and finally
reverse the other. However, for concentric cylinders the streamlines are concentric
circles; if we move one of the cylinders by angle $\theta$, a tracer particle will move
along a circle an angle that only depends on $\theta$. Then it is obvious
that the cumulative effect of moving one cylinder $\theta_1$, then the other $\theta_2$,
then the first $-\theta_1$, and the second $-\theta_2$, is to return the particle to its
original position: there is no geometric phase, and unmixing still occurs. But if we
modify the Heller--Taylor setup and offset the inner cylinder, we arrive at what is known
as journal--bearing flow. On introducing an eccentricity $\varepsilon$ between the
cylinders, this flow has a radial component. In the creeping-flow limit, the
Navier--Stokes equations for the journal--bearing flow reduce to a linear biharmonic one,
$\nabla^4\psi = 0 ,\label{bihar}$ for the stream function, $\psi$, and we may model this
system utilizing an analytical solution (see \cite{jeffery,ballal,finn} and the Materials and Methods
section for further details). If we now perform a parameter loop by the sequence of
rotations detailed above, we arrive back at our starting point from the point of view of
the positions of the two cylinders, so it is, perhaps, surprising that the fluid inside
does not return to its initial state. We illustrate the presence of this geometric phase
in Fig.~\ref{jbf_traj} in which an example of the trajectory of a fluid particle is shown
as the walls are driven through a nonreciprocal contractible loop. Journal--bearing flow
has been much studied in the past \cite{aref2,chaiken,ottino,tabor}, but never with
contractible loops so that this geometric effect was never emphasized. This minor
protocolary modification in a well established flow has, nonetheless, a substantial effect
on the fluid dynamics as we describe below.

We note that since a flow produced by a reciprocal cycle of the boundaries induces an
identity map for the positions of each fluid element at successive cycles, the problem of
mixing by nonreciprocal ones is closely related to the class of dynamical systems
constituted by perturbations of the identity. A fluid particle that at the beginning of the loop is in a position $\bar{x}=(x,z)$, reaches, at
the end of the same loop, a unique corresponding point $(x',z')$ which is a one-to-one
function $(x',z')=\textbf{G}[(x,z)]$ of the initial one. For homogeneous fluids,
$\textbf{G}$ must also be continuous and differentiable, whereas incompressibility implies
that $\textbf{G}$ preserves the area of any domain of points. In other words,
incompressible flow implies Hamiltonian dynamics for the fluid particles, and the map that
this dynamics induces in one loop is area preserving. For contractible zero-area loops the
map is simply the identity; each particle ends in the position in which it started. Hence,
a finite-area loop induces, in general, a finite deviation from the identity map and a
characteristic value of the geometrical phase gives an estimate for the extent of this
deviation. Since generically the geometric phase increases with the area of the loop (see
Fig.~\ref{line_evolution}(a)), for small loops the map is a small perturbation away from
the identity whereas loops of greater area induce larger deviations.

\begin{figure*}[h]
\begin{center}
\includegraphics[width=1.0\columnwidth]{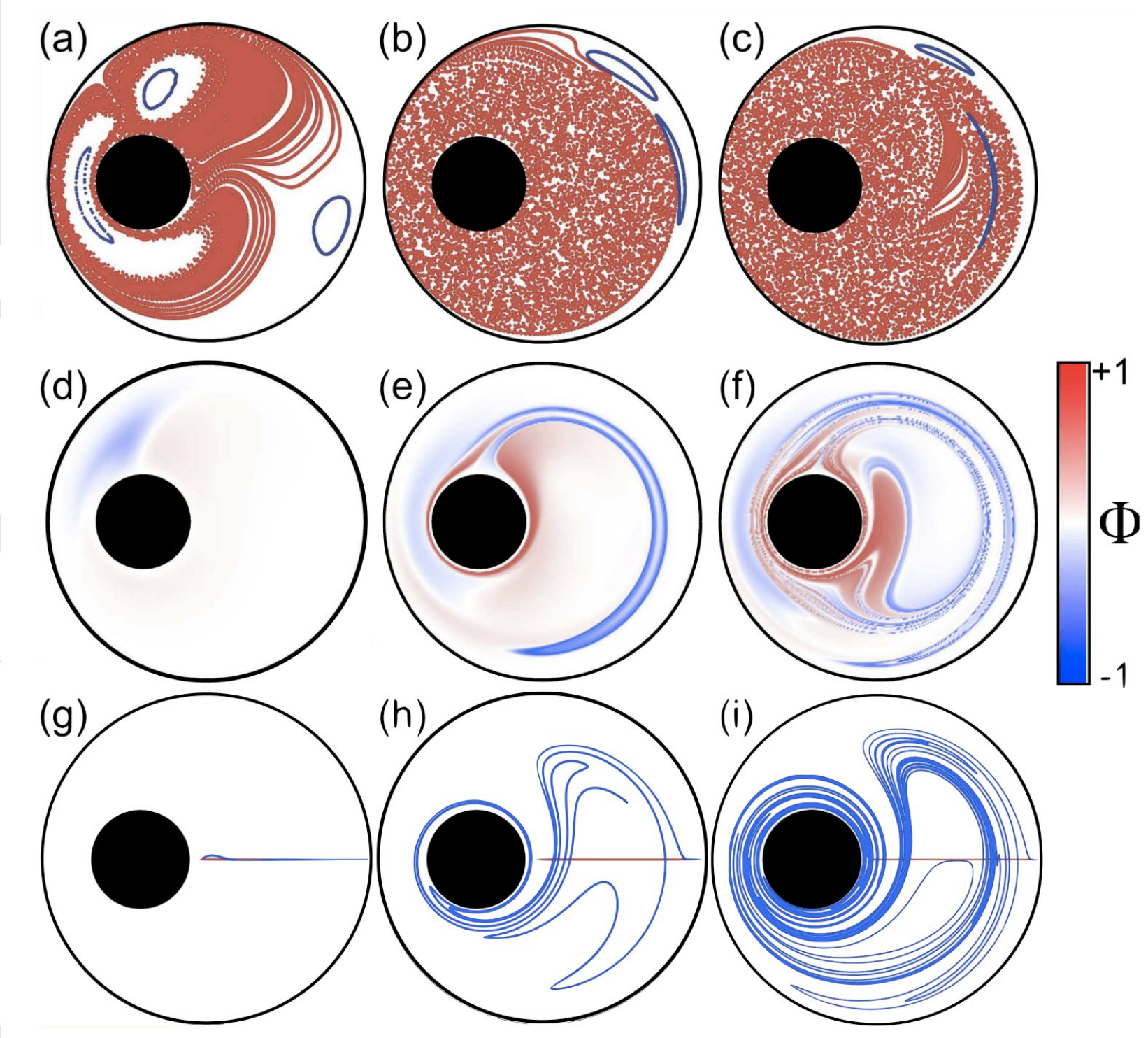}
\end{center}
\caption{{\bf Geometric mixing in the journal bearing flow.} (a--c) Poincar\'e maps
demonstrate geometric mixing for the journal--bearing flow for the same cylinder radii and
eccentricity as in Fig.~\ref{jbf_traj}. Chaotic trajectories are marked in red while
regular ones appear in blue. $10\,000$ iterations of the parameter loop are shown for: (a)
$\theta=\pi/2$ radians; (b) $\theta=2\pi$ and (c) $\theta=4\pi$. (d--f) The geometric
phase across the domain for the same parameters; the color scale denoting the phase at a
given point is given by the intensity of red, positive, and blue, negative. (g--i) The
evolution of a line segment --- initial segments in red; final segments in blue --- across
the widest gap between the cylinders after one cycle for the same parameters.}
\label{jb_phase}
\end{figure*}

\begin{figure*}[h]
\begin{center}
\includegraphics[width=1.0\columnwidth]{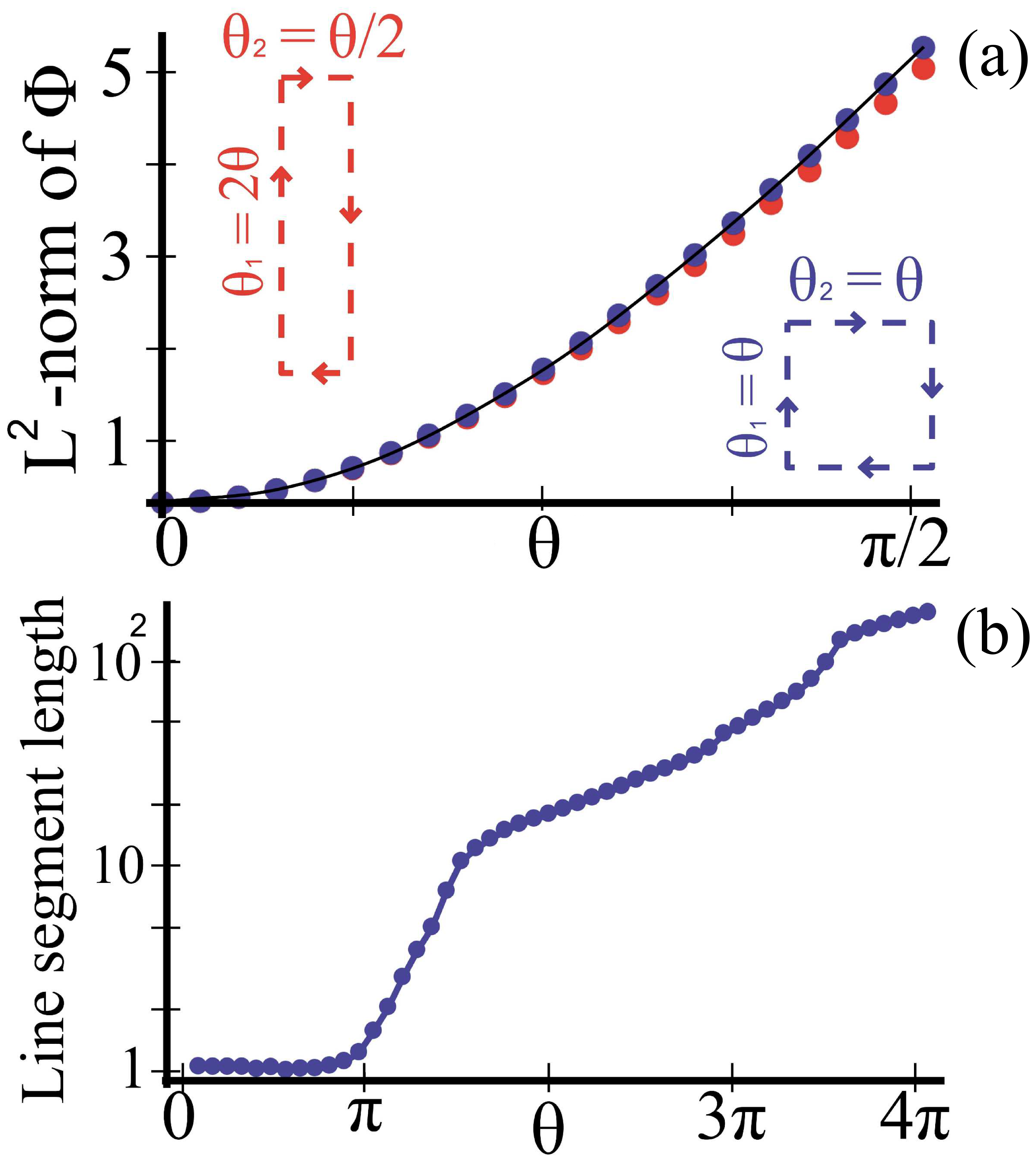}
\end{center}
\caption{{\bf Geometric phase correlates with the amount of stretching.} (a) The
$L^2$-norm of the $\Phi$ field grows quadratically with $\theta$ for loops with small
area. Two distinct loops with equal area are shown. (b) The final length of a line segment
as shown in Fig.~\ref{jb_phase} (g--i) plotted after one cycle for flows with different
rotation angles $\theta$.}
\label{line_evolution}
\end{figure*}

Let us now consider the long-term fluid dynamics elicited by a repeated realization of the
same contractible non-reciprocal loop that induces a given map. The dynamics is described
by the repeated iteration of this map that acts as the stroboscopic map of the
time-periodic Hamiltonian system constituted by the incompressible flow periodically
driven by the motion of the walls. For small loops, the map is a small perturbation of the
identity that can be thought of as the implementation of the Euler algorithm for a
putative continuous time dynamical system defined by this perturbation. Therefore, in 2D
we expect that the iterations of the map will closely follow the trajectories of this 2D
continuous system which is integrable. Therefore, fluid particles will mix very slowly in
space: this is, so to speak, mixing by "quasi-static" fluids. This is nicely illustrated
in Fig.~\ref{jb_phase}(a), where even for a square loop formed with values as large as
$\theta = \pi/2$ the positions of fluid particles after successive loops smoothly shift
along the closed curves that are the trajectories of the continuous dynamics. The
trajectories are composed of segments that nearly follow the integrable trajectories of a
2D flow (approximated as an Euler map) until it reaches the region of large phase, where
chaos and heteroclinic tangles occur. There the particle jumps into another
quasi-integrable trajectory, until it again reaches the region of large phase. In typical
Hamiltonian chaos (the standard map, for example) the map is not a perturbation of the
identity but a perturbation of a linear shear (i.e. with the canonical action-angle dynamical variables (I,$
\varphi$) following $I'=I$, $\varphi'=\varphi+I'$) for which reason this behavior is not normally
seen \cite{JoseSaletan,lichtenberg}. The structure of chaos in this class of dynamics has been greatly
overlooked in the literature, and the present research opens a new avenue to the
understanding of this associated problem.

As the geometric phase and the corresponding perturbation from the identity map increase,
the former argument begins to fail \cite{rkpaper}. A more chaotic 2D-area preserving map
emerges and with it the corresponding space-filling fully chaotic trajectories. The KAM
islands typically become smaller and smaller as the characteristic values of geometric
phase increase. As we see in Fig.~\ref{jb_phase}(b) for $\theta=2\pi$ radians, and even
more so in \ref{jb_phase}(c) for $\theta=4\pi$ radians, after $10\,000$ cycles the fluid
particle has covered most of the area available to it between the two cylinders. This is
fluid mixing induced entirely by a geometric phase; we may call it geometric mixing.
Geometric mixing therefore creates chaotic advection~\cite{ottino}, as does the classical
journal--bearing protocol.

In Fig.~\ref{jb_phase}(d--f) we show the corresponding distributions of the geometric
phase over the domain. The value of the geometric phase at a given initial position,
obtained in terms of the final angle minus the initial angle in bipolar coordinates (see the Materials and Methods
section for further details) after
one iteration, $\Phi=\xi_f-\xi_i$, is plotted on a color scale of intensities of red
(positive) and blue (negative). Note that the phase goes to zero at the walls, as it must,
but varies strongly across the domain. In particular, for parameters of $\theta=2\pi$
radians (Fig.~\ref{jb_phase}(e)), we see the development of a tongue of high values of the
geometric phase in one sense interpenetrating a region of high values of the phase in the
opposite sense. The trajectory plotted in Fig.~\ref{jbf_traj} shows the origin of the
tongue; fluid particles that are advected to the vicinity of the inner cylinder by the
first $\theta_1$ step are then advected to a significantly different value of $r$ by the
inner cylinder. As a result, the fluid particle is located on a completely different
streamline from the first step when the outer cylinder starts rotating backwards. As may
be expected, for smaller parameter values this tongue is absent (Fig.~\ref{jb_phase}(d)).
At even higher values of $\theta$, on the other hand, (Fig.~\ref{jb_phase}(f)) the tongue
wraps twice round in a highly complex fashion. In Fig.~\ref{jb_phase}(g--i) we show by
plotting the evolution of a line of initial conditions how the geometric phase is related
with the dynamical structures in the flow. Fig.~\ref{jb_phase}(g), for $\theta=\pi/2$
radians, shows that when this tongue is absent, the line segment hardly evolves; the flow
is almost reversible. The line segments for Fig.~\ref{jb_phase}(h), and (i), for $\theta=2
\pi$ and $4\pi$ radians, on the other hand, show a great deal of stretching induced by
this tongue of large geometric phase. To demonstrate this effect of the geometric phase on
the flow in more detail, in Fig.~\ref{line_evolution}(b) we plot the length of the line
segment after a single cycle against the rotation angle. A notable aspect of this plot is
that it displays plateaux separated by periods of rapid growth. A comparison with
Fig.~\ref{jb_phase}(d--f) shows that it is the penetration of the tongue of large values
of the geometric phase across this line segment that induces stretching. The tongue
penetrates a first time before $\theta=2\pi$, and then a second time before $\theta=4\pi$,
so producing two jumps; between these jumps the evolution of the line segment is much
slower. For a given energy cost, which scales with the total unsigned displacement of the
walls, geometric mixing is therefore more efficient for a large value of $\theta$.

The journal--bearing flow is just one member of a class of flows that display geometric
mixing. In open flows, one has instances such as the well-known Purcell swimmer that can
be seen as operating through a geometric phase. Another closed flow that was studied early
on in chaotic advection is the rectangular cavity flow, in which one or more of the walls
of a fluid filled rectangular container can move, being set up as conveyor belts
\cite{ottino2,ottino}. As in the case of the former studies of the journal--bearing, these
mixing protocols imply a cumulative relative displacement of the container walls. However,
in the same way as in the journal--bearing case one can introduce a geometric phase by
returning all the walls to their initial relative positions after a loop in the
parameters. More generally, one can conceive of flows induced by a container in which the
walls do not move as rigid bodies, but instead can deform longitudinally and/or
tangentially along a nonreciprocal cycle in order to produce efficient mixing. For
example, one might consider the case of an elastic bag containing a fluid and subject to
the action of a periodic squeezing--distention sequence around one of its sections with a
compensating distention--squeezing action around another. This cycle would clearly induce
a reciprocating flow unfit for efficient mixing, but again a geometric phase could be made
to exist if this spatially stationary configuration were replaced by one that propagates
along the bag axis.


\subsection*{Peristaltic mixing}

The stomach is a biological instance of such a cavity flow \cite{Pal04,schulze}. The human stomach is a strong muscular receptacle between the oesophagus and the small intestine. It is not just a storage chamber for food, but also a mixer where the chyme is prepared. The human stomach has a volume $L^3$ of some 330~mL, while the viscosity $\mu$ of the chyme is of order 1~Pa\,s, its density is $\rho\approx10^3$~kg\,m$^{-3}$, and the maximum flow velocities $V$ observed are in the range 2.5--7.5~mm\,s$^{-1}$ \cite{Pal04}. From these data we may estimate the Reynolds number $Re=\rho V L/\mu$ to lie in the range 0.2--0.5. Thus we may conclude that in the human stomach fluid inertia has only limited importance, and in any smaller animal it will be inappreciable. We note that previous work on gastric mixing have mostly considered the case of inertial contributions \cite{Dillard07, Kumar07, Pal04} for which the dynamical constraints discussed herein do not apply. 

The gastric mixing is brought about by peristaltic waves --- transverse traveling waves of contraction --- that propagate along the stomach walls at some 2.5~mm\,s$^{-1}$. They are initiated approximately every 20~s, and take some 60~s to pass the length of the stomach, so 2--3 waves are present at one time, while on average the stomach width as the wave passes is 0.6 times its normal width \cite{Pal04, schulze}. We thus have their velocity $c=2.5$~mm\,s$^{-1}$, frequency $\omega=0.05$~Hz, and thence wavelength $\lambda=c/\omega=5$~cm, and their amplitude $b=1/2\times0.6L\approx 2$~cm. These waves force the stomach through a nonreciprocal loop in the space of shapes, as a result of which geometric mixing is expected. One can give a rough estimate of the size of the expected geometric phase by taking advantage of results obtained for another geometric phase problem: that of low-Reynolds-number microorganisms swimming. Many bacteria swim by deforming their bodies in the same way as the peristaltic waves of the stomach and their speed has been well estimated by modeling such deformations as plane waves \cite{koiller}. Similar calculations for the stomach render the flow velocity induced by the peristaltic waves $V=\pi c (b/\lambda)^2$, which comes out at approximately 1~mm\,s$^{-1}$, from where a displacement of about 6 cm per peristaltic cycle is expected or, considering a circular stomach of radius L, a geometric phase of the order of $2$~radians.

\begin{figure*}[h]
\begin{center}
\includegraphics[width=1.0\columnwidth]{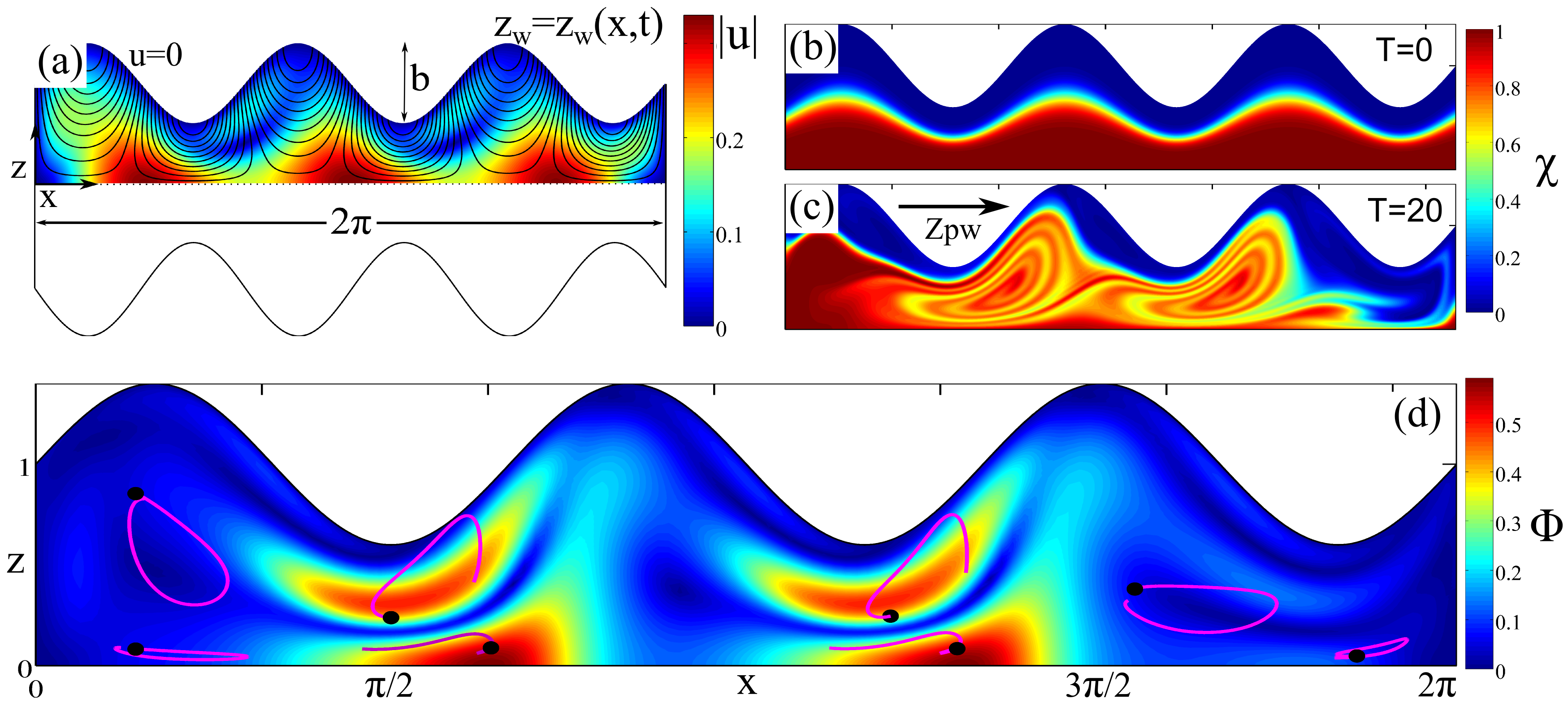}
\end{center}
\caption{{\bf Peristaltic mixing is generated by a geometric phase.} (a) The minimal
geometry of a model for the stomach. A peristaltic wave propagates along the upper and
lower boundaries of a closed cavity $z_w(x,t)=1+b\sin\left(kx-\omega t\right)$ of aspect
ratio $\pi$. (b) Contours of concentration of a passive scalar whose initial spatial
distribution at $t=0$ is given by a blurred step with $\delta=0.1$. The temporal evolution
of the spatial concentration is obtained integrating the advection--diffusion equation for
$Pe=15\times 10^3$. (c) The spatial concentration after 20 peristaltic cycles. (d) The
geometric phase of the system. Pink solid lines show some representative trajectories with
initial conditions marked by thick black dots.}
\label{stomach}
\end{figure*}

To show the effects of this phase, we have constructed a minimal model of the stomach undergoing peristalsis, as sketched in Fig.~\ref{stomach}(a) and further detailed in the Materials and Methods section. We have intentionally reduced the geometric, dynamic, and functional complexity of the stomach and model a 2D section of a tubular stomach of uniform radius, with sealed pyloric and esophageal valves, to focus on the role peristaltic contractions may play in mixing within the enclosed inertialess cavity. Similarly, we have treated the chyme as a Newtonian fluid, leaving the complexities associated with viscoelasticity for future work, as the existence or not of geometric mixing in the stomach is independent of the viscoelastic properties of the chyme. A similar model was used in \cite{Shapiro1969, Jaffrin1971} to assess transport by peristaltic pumping in infinite slender tubes. In our model a peristaltic wave deforms the upper and lower boundaries of a symmetric cavity of aspect ratio $\pi$ according to $z_w(x,t)=1+b\sin\left(kx-\omega t\right)$ in the $(x,z)$-coordinate system. The flow within the cavity is obtained by integrating the Stokes equations for the velocity field $(u,v)$ with the corresponding boundary conditions for the peristaltic wave at the upper boundary, $u=0$ and $v=\partial z_w/\partial t$ at $z=z_w(x,t)$, and symmetry boundary conditions at $z=0$. Lateral walls deform vertically to match the vertical velocity of the peristaltic wave at $x=0$ and $x=2\pi$. In Fig.~\ref{stomach}(a) black solid lines represent the streamlines of the induced fluid motion within the cavity due to the peristaltic wave. The contour plot corresponds to the time-averaged velocity over one full peristaltic cycle. Areas of maximum average velocity are close to the axis of symmetry, whereas near the wall the average velocity is zero and no average motion is produced.


We consider the mixing of a passive scalar $\chi(\bar{x},t)$ whose initial spatial distribution at $t=0$ is given by the blurred step ($\chi(\bar{x},t=0)=1+\tanh[(z/z_w-1/2)/\delta])$,  as represented in the contour map of Fig.~\ref{stomach}(b). The temporal evolution of this spatial concentration is obtained integrating the advection-diffusion equation for a characteristic P\'eclet number, $Pe=c\lambda/D_{chyme}$ representative of the mixing process within the stomach. As the characteristic diffusivity of the chyme is, at most, of order of the molecular diffusion of large macromolecules $D_{chyme}\leq10^{-6} cm^2/s$, $Pe\gg 1$ and advective contributions dominate the mixing process.  Fig.~\ref{stomach} (c) represents the spatial concentration of the passive scalar $\chi$ after 20 peristaltic cycles (i.e. after a rescaled time $T=t/T^*=20$, where $T^*$ represents the cycle period) for $Pe=15\times 10^3$. The flow induced by peristalsis accumulates a finite geometric phase after each cycle, fluid elements are stretched and folded and, as a consequence, thin filaments are formed that facilitate mixing within the cavity.

We obtain the geometric phase by integrating the trajectory of passive scalars over one full cycle, with uniformly distributed initial conditions in the domain $[0,2\pi]\times[0,z_w(x,0)]$. The Euclidean distance between the initial and final position after one cycle gives an estimate of the geometric phase. Contours in Fig.~\ref{stomach}(d) represent the geometric phase of the system. It can be seen that maximum displacements are observed in the central region of the cavity where filaments are created. Note that regions in Fig.~\ref{stomach}(d) with small displacements correspond to regions that remain unmixed in Fig.~\ref{stomach}(c). Thus, and despite the uniform radius of the cavity in our minimal geometric model, mixing is not spatially uniform. Regions in the central part of the cavity form thin filaments that enhance mixing, whereas regions close to the lateral and to the upper walls remain almost unmixed after 20 cycles. Even further inhomogeneities are expected for more faithful geometries, with changing average wall diameter \cite{Kwiatek}, specific timing of the opening and closing of the pylorus with peristalsis \cite{Indireshlumar} and interactions between the fundic/cardiac region of the stomach \cite{Imai2013}, all of which are known feature for mixing within the stomach \cite{Pal07}.

\begin{figure*}[h]
\begin{center}
\includegraphics[width=1.0\columnwidth]{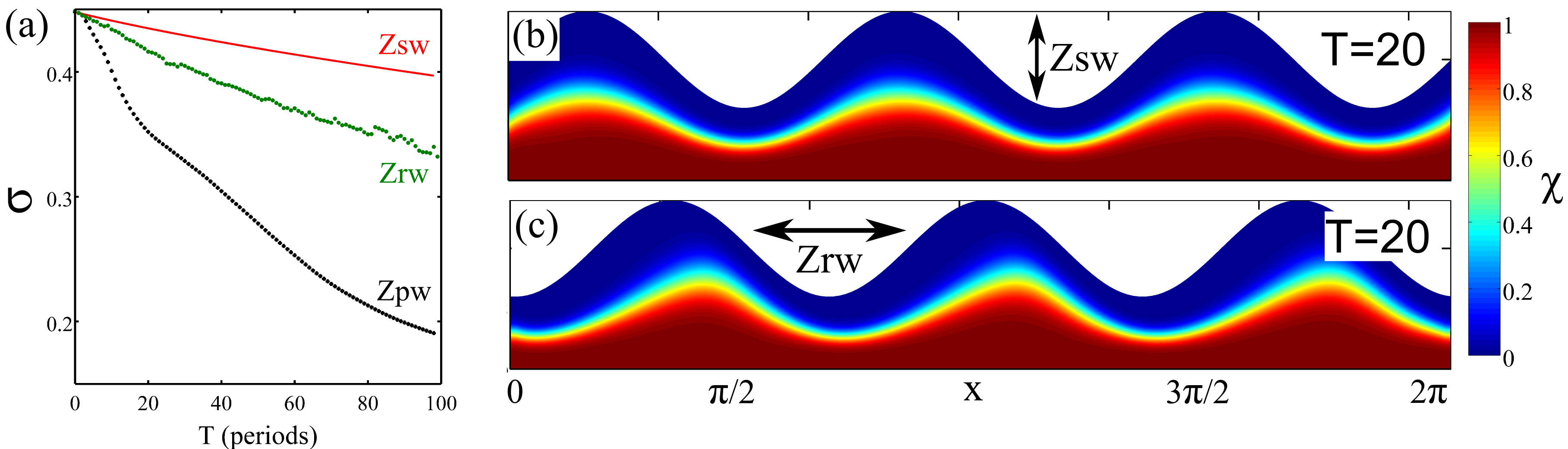}
\end{center}
\caption{{\bf Mixing quality depends on the accumulated geometric phase.} (a) The time
evolution of the degree of mixing quantified by the standard deviation of the
concentration field in the domain. The black dotted line corresponds to the peristaltic
wave; the red solid line to the standing wave and the green dotted line to the random
wave. (b) and (c) show contours of concentration of $\chi$ after the same integration
time, equivalent to 20 peristaltic cycles, shown in fig.~\ref{stomach} (c) for the case of
a stationary and random wave, respectively.}
\label{standing}
\end{figure*}

Stomach contractions that correspond to a standing wave are akin to a zero-area reciprocal loop. As we anticipated for the journal-bearing case, reciprocal loops induce flow which does not generate any mixing. This is shown in Fig.~\ref{standing}(b) where the concentration field after 20 cycles of the boundaries deforming as a standing wave is depicted. Since the induced geometric phase is null, mixing is only controlled by (slow) diffusion.

The importance of the geometric mixing in the stomach may be appreciated by reference to instances in which it is disrupted. The stomach is like the heart, with electrical activity from a pacemaker region stimulating oscillations; in this case being traveling waves of peristalsis. If this system is not functioning correctly, there can be gastroparesis or gastric fibrillation \cite{lin,rinaldi}, in which the peristaltic waves become disordered. We have generated such disordered deformations by interspersing peristaltic waves whose propagation velocities $c$ are chosen randomly from a uniform distribution of zero mean. The scalar field $\chi$ remains almost unmixed compared to the peristaltic case after an equivalent integration time, with mixing mostly controlled again by slow diffusion (Fig.~\ref{standing}(c)). Thus, in our terms, there is poor mixing or no mixing in gastroparesis because there is not a loop around the space of forms, so no average geometric phase, and instead random peristaltic waves induce only mixing and unmixing.

To compare the degree of mixing in the three cases considered herein (peristalsis (pw), stationary (sw) and random (rw) waves), we calculate for each cycle the variance of the spatial concentration field \cite{Stroock02,Thiffeault2012}, $\sigma=\left\langle(\chi-\langle\chi\rangle)^2\right\rangle^{1/2}$, where $\langle\rangle$ denotes the spatial average. Fig.~\ref{standing} (a) represents the evolution of $\sigma$ with the number of cycles. It reveals the higher mixing efficiency realized in peristalsis by geometric mixing.

\section*{Discussion}

In summary, we have introduced the concept of geometric mixing in which mixing arises as a consequence of a geometric phase induced by a contractible  non-reciprocal cycle in the parameters defining the shape of the container. It turns out that the mixing efficiency estimated from the stretching of material lines is roughly proportional to the geometric phase. Mixing in the corresponding flows can be also considered as the result of chaos arising in the mapping describing the motion of fluid elements during one  cycle. When the cycle is reciprocal, this map is the identity and a small departure from reciprocity corresponds to a small departure from the identity map. The chaotic properties of maps neighboring the identity have been poorly studied in the past. They also arise in the quite different context of numerical integration methods of ordinary differential equations in the limit where the step size tends to zero \cite{rkpaper}. Our results are therefore also relevant to the characterization of chaos in this class of systems. Lastly, we have shown that such a geometric phase ---  the ``belly phase'' \cite{berry} ---  may be found in the stomachs of animals where $Re<1$.

\section*{Supporting Information}

\subsection*{Journal bearing flow}
\begin{figure*}[h]
\begin{center}
\includegraphics[width=1.0\columnwidth]{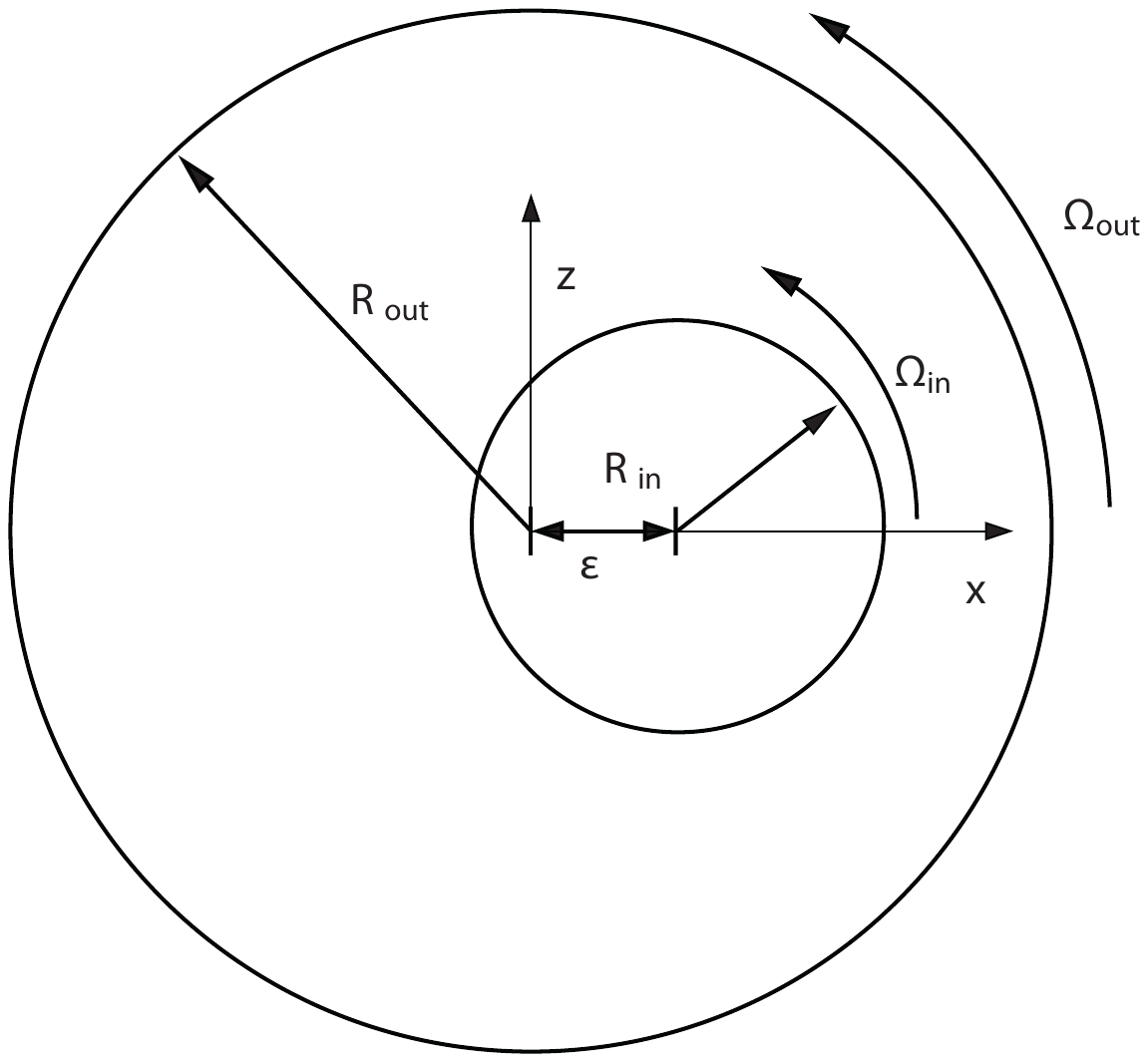}
\end{center}
\caption{Sketch of the configuration of the journal-bearing flow. See text for details.}
\label{jbf_sketch}
\end{figure*}

The journal-bearing flow has been widely employed to study the process of mixing in
laminar flows. Fig.~\ref{jbf_sketch} shows a sketch of the configuration studied herein.
The outer cylinder of radius $R_{out}$ rotates with an angular velocity
$\Omega_{out}$, whereas the inner cylinder of radius $R_{in}$ rotates with an angular
velocity $ \Omega_{in}$. The eccentricity of the inner cylinder is given by
$\varepsilon$. In the limit where viscous forces are negligible, the resulting flow is
obtained by integrating the biharmonic equation for the stream function $\nabla^4\psi=0$
with the corresponding boundary conditions at the walls of the cylinders.

Because of the linearity of the problem the solution for the stream
function can be written as
\begin{equation}
\psi=\Omega_{out}\psi_{out}+\Omega_{in}\psi_{in},
\label{jbf_eq}
\end{equation}  
where $\psi_{out}$ is the solution for the stream function of the flow induced by the
outer cylinder, whereas $\psi_{in}$ corresponds to the solution of the stream function of
the flow induced by the inner cylinder. The angle covered by a cylinder during one cycle depends on its angular velocity according to
\begin{equation}
\Theta_i=\int_{0}^{T^{*}}\Omega_{i}(t)\mathrm{d}t,
\end{equation}
where the subindex $i$ denotes the outer or inner cylinder and $T^*$ represents the period of the cycle. Since, in the simulations considered in this paper the angular velocity of the cylinders is constant, $\Theta_i=T^*\Omega_i$. This flow admits
an exact solution \cite{ballal} for the stream function when the problem is written in bipolar coordinates, $(\xi,\eta)$. The cartesian coordinates $(x,z)$ can be recovered according to
\begin{equation}
x=-b\frac{\sinh\xi}{\cosh \xi-\cos\eta},\; z=b\frac{\sin\eta}{\cosh\xi-\cos\eta},
\label{bipolar_coord}
\end{equation}
where 
\begin{equation}
b=\frac{1}{2\varepsilon}[(R_{in}^2+R_{out}^2-\varepsilon^2)^2-4R_{in}^2R_{out}^2]^{1/2}.
\end{equation}
Following \cite{ballal} the solution for the inner and outer stream functions is given by
\begin{equation}
\psi=H\phi,\; \phi=F_0(\xi)+F_1(\xi)\cos\eta,
\end{equation}
where $H=b/(c^2+s^2)^{1/2}$, with $s=\sin\xi\sin\eta$ and $c=cos\xi\cos\eta-1$. Moreover, 
\begin{align}
F_0(\xi)&=(A_0+C_0\xi)\cosh\xi+(B_0+D_0\xi)\sinh\xi,\\
F_1(\xi)&=A_1\cosh 2\xi+B_1\sinh 2\xi+C_1\xi+D_1, 
\end{align}
and
\begin{align}
(A_0,B_0,C_0,D_0)=&(f_1,f_3,f_5,f_7)\Omega_{out}R_{out}+\\ \nonumber
&(f_2,f_4,f_6,f_8)\Omega_{in}R_{in}\\
(A_1,B_1,C_1,D_1)=&(f_9,f_{11}-f_5,f_{13})\Omega_{out}R_{out}+\\ \nonumber
&(f_{10},f_{12}-f_6,f_{14})\Omega_{in}R_{in}
\end{align}
with
\begin{flalign*}
&f_1=\frac{1}{\Delta}\left(\frac{\bar{\Delta}}{\Delta^*}h_1 h_7+h_3\right),\
f_2=\frac{1}{\Delta}\left(\frac{\bar{\Delta}}{\Delta^*}h_2 h_7+h_4\right),&\\
&f_3=\frac{1}{\Delta}\left(\frac{\bar{\Delta}}{\Delta^*}h_1 h_8+h_5\right),\
f_4=\frac{1}{\Delta}\left(\frac{\bar{\Delta}}{\Delta^*}h_2 h_8+h_6\right),&\\
&\Delta^* f_5/h_1=\Delta^* f_6/h_2=\cosh(\xi_{out}-\xi_{in}),&\\
&\Delta^* f_7/\sinh\xi_{in}=\Delta^* f_8/\sinh\xi_{out}=-\sinh^2(\xi_{out}-\xi_{in}),&\\
&2\Delta^* f_9/h_1=2\Delta^* f_{10}/h_2=-\sinh(\xi_{out}+\xi_{in}),&\\
&2\Delta^* f_{11}/h_1=2\Delta^* f_{12}/h_2=\cosh(\xi_{out}+\xi_{in}),&\\
&2\Delta^* f_{13}/h_1=2\Delta^* f_{14}/h_2=\sinh(\xi_{out}-\xi_{in})+2\xi_{in}\cosh(\xi_{out}-\xi_{in}),&
\end{flalign*}
where $\xi_{in}$ and $\xi_{out}$ represent the surfaces of the inner and outer cylinders, respectively, and $\Delta$, $\bar{\Delta}$, $\Delta^*$, $h_1$, $h_2$, ..., $h_8$ are given by
\begin{flalign*}
&\Delta=(\xi_{out}-\xi_{in})^2-\sinh^2(\xi_{out}-\xi_{in}),& \\
&\bar{\Delta}=(\xi_{out}-\xi_{in})\cosh(\xi_{out}-\xi_{in})-\sinh(\xi_{out}-\xi_{in}),& \\ 
&\Delta^*=\sinh(\xi_{out}-\xi_{in}) [2\sinh\xi_{out}\sinh\xi_{in}\sinh(\xi_{out}-\xi_{in})-&\\ 
&(\xi_{out}-\xi_{in})(\sinh^2\xi_{out}+\sinh^2\xi_{in})],&\\
&h_1=(\xi_{out}-\xi_{in})\sinh\xi_{out}-\sinh\xi_{in}\sinh(\xi_{out}-\xi_{in}),&\\ 
&h_2=-(\xi_{out}-\xi_{in})\sinh\xi_{in}+\sinh\xi_{out}\sinh(\xi_{out}-\xi_{in}),&\\ 
&h_3=\xi_{out}\sinh\xi_{in}\sinh(\xi_{out}-\xi_{in})-\xi_{in}(\xi_{out}-\xi_{in})\sinh\xi_{out},&\\ 
&h_4=-\xi_{in}\sinh\xi_{out}\sinh(\xi_{out}-\xi_{in})+\xi_{out}(\xi_{out}-\xi_{in})\sinh\xi_{in},&\\ 
&h_5=-\xi_{out}\cosh\xi_{in}\sinh(\xi_{out}-\xi_{in})+\xi_{in}(\xi_{out}-\xi_{in})\cosh\xi_{out},&\\ 
&h_6=\xi_{in}\cosh\xi_{out}\sinh(\xi_{out}-\xi_{in})-\xi_{out}(\xi_{out}-\xi_{in})\cosh\xi_{in},&\\ 
&h_7=\sinh\xi_{in}\cosh\xi_{out}\sinh(\xi_{out}-\xi_{in})+1/2\xi_{out}\sinh 2\xi_{in}-&\\ 
&1/2\xi_{in}\sinh2\xi_{out}-\xi_{in}(\xi_{out}-\xi_{in}),\\ 
&h_8=-\cosh\xi_{out}\cosh\xi_{in}\sinh(\xi_{out}-\xi_{in})+\xi_{in}\cosh^2\xi_{out}-&\\
&\xi_{out}\cosh^2\xi_{in}&.
\end{flalign*}

Once the flow is evaluated, the trajectories of the particles were obtained integrating
\begin{align}
\frac{d\xi}{dt}&=\frac{1}{H^2}\frac{\partial\psi}{\partial\eta},\label{particle_motion1}\\
\frac{d\eta}{dt}&=-\frac{1}{H^2}\frac{\partial\psi}{\partial\xi},
\label{particle_motion2}
\end{align}
The integration of \eqref{particle_motion1} and \eqref{particle_motion2} was carried out with a fourth order Runge-Kutta scheme. After one complete cycle, both cylinders end at their initial position, while particles departing from an initial position $(\xi_i,\eta_i)$ are located at $(\xi_f,\eta_f)$. A stroboscopic map was constructed from the iterative application of the described loops which characterizes qualitatively the dynamics of particles in the journal-bearing flow. Final positions of the particles after each full iteration were plotted to distinguish between particles that follow smooth integrable trajectories from particles following space-filling fully chaotic trajectories (see Fig.~2 of the main text). We compute the geometric phase as the travelled distance measure as the difference in the bipolar angle $\Phi=\xi_f-\xi_i$. 

We also evaluate the mixing efficiency by computing the length by which an initial fluid segment is stretched as a function of the number of mixing cycles. For this, we densely sample the initial fluid segment with $N$ particles, $\mathbf{r_i}(0), i = 1, ..., N$, and track the motion of each one by the method described above after $T$ iterations of the mixing protocol, obtaining the final position of each particle, $\mathbf{r_i}(n=T), i = 1, ..., N$. We then compute the total length of the final segment by adding the distances between adjacent points as $\sum{\left|r_i(T)-r_{i-1}(T)\right|}_{i=2}^N$. The true segment length is given by the limit of this expression for $N \rightarrow \infty$. In practice, we make sure the segment was sampled densely enough by computing the final segment length using $N$ and $2 N$ particles, with $N$ sufficiently large such that both computations give the same result to within a fraction of the initial segment length.

\subsection*{Cavity flow}

\begin{figure*}[h]
\begin{center}
\includegraphics[width=1.0\columnwidth]{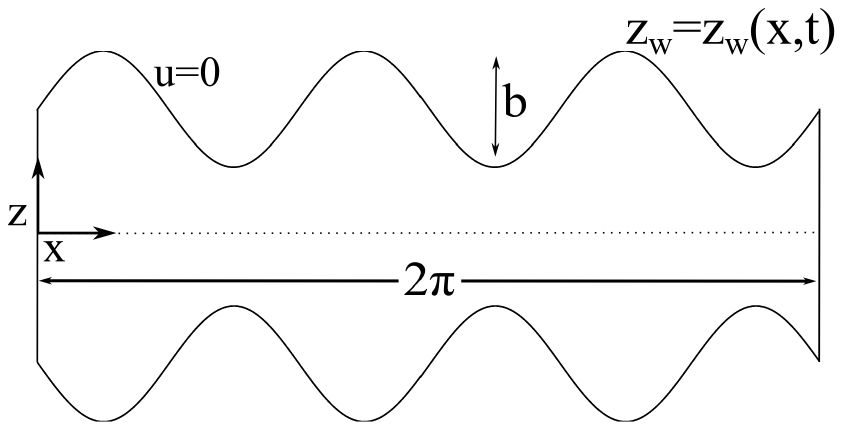}
\end{center}
\caption{Sketch of the configuration of the cavity flow. See text for details}
\label{cavity_sketch}
\end{figure*}

We studied the flow within a deformable symmetric cavity, sketched in Fig.~\ref{cavity_sketch}. A peristaltic wave propagates along the upper and lower walls of a
two dimensional cavity of length $2\pi$, according to $z_w(x,t)=1+b\sin(kx-\omega t)$,
where $b$ is the amplitude of the wave, $k$ its wavenumber and $\omega$ its angular
velocity. This two-dimensional configuration corresponds to a section of a cylindrically
symmetric cavity around the $x$-axis (i.e. a symmetrically deforming tube). All lengths are
normalized by the undeformed tube radius and we keep the name of the non-dimensional variables the same for simplicity. We set the cavity length-to-width aspect ratio to $\pi$ in all simulations.

\paragraph*{Numerical solution to the cavity flow}

The propagation of the waves along the no-slip boundaries induces a flow within the cavity that,
in the limit where viscous forces are dominant, is obtained by integrating the
two-dimensional Stokes equations
\begin{align}
&\frac{\partial u}{\partial x}+\frac{\partial v}{\partial z}=0,\label{cont_eq}\\
&\frac{\partial^2 u}{\partial x^2}+\frac{\partial^2 u}{\partial z^2}=0,\label{mom_eq_x}\\
&\frac{\partial^2 v}{\partial x^2}+\frac{\partial^2 v}{\partial z^2}=0,\label{mom_eq_z}
\end{align}
with the corresponding boundary conditions: $u=0$ and $v=\partial z_{w}/ \partial t$ at
the upper wall (i.e. $z=z_w$) and $\partial u/\partial z=0$ and $v=0$ at $z=0$, to impose
symmetry conditions. Lateral walls deform vertically (with a linear vertical displacement
distribution) to match the vertical velocity of the peristaltic wave: $u=0$ and
$v=z\partial z_{w}/ \partial t|_{x=0}$ at $x = 0$, and $u=0$ and $v=z\partial z_{w}/
\partial t|_{x=2\pi}$ at $x=2\pi$. Initial conditions correspond to the fluid
at rest.

To facilitate the numerical integration of the problem we employed the vorticity-stream
function formulation \cite{Leal07}. Thus, the problem reduces to that of integrating
\begin{align}
\nabla^2 \psi&=-\bar{\omega},\label{stream_eq}\\
\nabla^2 \bar{\omega}&=0,\label{vorticity_eq}
\end{align} 
where $\bar{\omega}=-(\partial u/\partial z-\partial v/\partial x)$ is the
vorticity of the flow and $\psi$ its stream function. Accordingly, the boundary conditions
given above for the velocity field have to be expressed in terms of $\bar{\omega}$ and $\psi$.

We note that the computational domain changes as the moving boundaries evolve in time.
However, this domain was transformed into a fixed domain by means of a mesh transformation
\cite{Anderson84} with a new computational vertical coordinate, $Z=z/z_w(x,t)$. Equations
\eqref{stream_eq} and \eqref{vorticity_eq} were rewritten in terms of the new variable $Z$
and were discretized with a second-order accurate finite difference scheme. The coupled
system of equations \eqref{stream_eq} and \eqref{vorticity_eq} was then solved with a
line-by-line Thomas algorithm \cite{Anderson84} at each time step. The iterative process
to solve \eqref{stream_eq} and \eqref{vorticity_eq} starts with an initial guess for the
vorticity and stream function fields which corresponded with the values of the previous
time step. This system of equations was solved iteratively until convergence was achieved.
The flow in a squared cavity studied in \cite{Ghia82} was used as a check for our
numerical scheme.

We integrate numerically equations \eqref{stream_eq} and \eqref{vorticity_eq} for a fixed
set of parameters with non-dimensional values consistent with the available experimental data for a human
stomach. We consider peristaltic waves with $b=0.4$, $k=3$ and $\omega=3$. The flow
created by a standing wave was obtained by superposing two peristaltic waves with opposite
velocities: the first moving from left to right with $\omega=3$ whereas the second
propagates from right to left with $\omega=-3$. As the resulting flows are time periodic
in these two cases, the integration only needs to be carried out for a single wave period,
$t=T^*=2\pi/\omega$, saving computational time.

To generate a random wave, we choose the angular velocities, $\omega_i$, sampling them
randomly at each period $T$ from a continuous uniform distribution with zero mean in the
interval $[-3,3]$, while keeping both the wavenumber and the amplitude fixed to the same
values used above. Thus, values of $\omega$ can be either positive or negative,
corresponding to waves propagating respectively from left to right or viceversa with
random angular velocity. Since the resulting flow is not periodic in this case, the
numerical integration was carried out for a prescribed integration time equivalent to 100
periods of the peristaltic wave, which facilitates the comparison among the different
scenarios. To ensure a continuous evolution of the waveform at the end of each period
(i.e. coincident with a change in $\omega$), we shift the new waveform along the x-axis
(i.e. we introduce a phase shift) to match the initial condition for the succeeding period.

\paragraph*{Mixing within the cavity}

To analyze mixing within the cavity we integrated the two-dimensional
advection-diffusion equation for a passive scalar $\chi(\bar{x},t)$:
\begin{equation}
\frac{\partial \chi}{\partial t}+u\frac{\partial \chi}{\partial x}+v\frac{\partial \chi}{\partial z}
-\frac{1}{Pe}\left(\frac{\partial^2 \chi}{\partial x^2}+\frac{\partial^2 \chi}{\partial z^2}\right)=0,
\label{ad-diff_eq}
\end{equation}
that is initially distributed in a blurred step ($\chi(\bar{x},t=0)=1+\tanh[(z/z_w-1/2)/
\delta])$), with $\delta=0.1$ its characteristic thickness (see Fig.~4(b) of the main text). 

The advection-diffusion process is characterized by a non-dimensional P\'eclet number,
$Pe=c\lambda/ D_{chyme}$, which in our simulations was fixed to $Pe=15\times10^3$ since in
the case considered here advection largely dominates over diffusion. Equation
\eqref{ad-diff_eq} was integrated with no-flux boundary conditions
$\bar{u}\chi-1/Pe\nabla\chi=0$ at the boundaries of the computational domain, together
with the symmetry boundary condition at $z=0$. Since the velocity field was provided by
the integration of \eqref{stream_eq}-\eqref{vorticity_eq}, we used the same transformation
of the domain and the same spatial discretization described above for the numerical
integration of \eqref{ad-diff_eq} together with a Crank-Nicholson scheme for the temporal
evolution.

To compare the degree of mixing in the three cases considered herein (peristalsis (pw),
stationary (sw) and random (rw) waves), we calculate for each cycle the variance of the
spatial concentration field \cite{Stroock02,Thiffeault2012},
$\sigma=\left\langle(\chi-\langle\chi\rangle)^2\right\rangle^{1/2}$, where
$\langle\rangle$ denotes the spatial average. Fig.~5(a) of the main text shows the
evolution of $\sigma$ with the number of cycles for the three cases we have considered.

Finally, to determine the geometric phase induced by the peristaltic wave we calculated
the trajectories of uniformly distributed passive particles within the cavity for one
compete period of the peristaltic wave. The trajectories were obtained integrating
$dx/dt=u,dz/dt=v$ with a fourth order Runge- Kutta scheme. As $(u,v)$ were evaluated at
each time step on the mesh employed to integrate \eqref{stream_eq}
and\eqref{vorticity_eq}, we employed a 2D linear interpolation to obtain their values at
the particles position. The distance from the position of the particle at the end of one
period to its initial position was used to determine the spatial distribution of the
geometric phase, $\Phi$, represented in Fig. 4(d) of the main text.

\section*{Acknowlegments}
We acknowledge the financial support of the grants FIS2010-22322-C02-01 and -02 from MICINN and from the subprograma Ram\'on y Cajal (I.T.).

\bibliography{PLOSonebib.bib}

\begin{thebibliography}{10}
\bibitem{berry}
{We are indebted to Michael Berry for this coinage};.
\end{thebibliography}

\pagebreak


\end{document}